\documentclass[preprint,nofootinbib,aps,superscriptaddress,floatfix]{revtex4}

\usepackage{graphicx}
\usepackage{epsfig}
\usepackage{dcolumn}
\usepackage{relsize}

\newif\ifpdf
\ifx\pdfoutput\undefined
\pdffalse 
\else
\pdfoutput=1 
\pdftrue
\fi

\def\OMIT#1{{}}
\def\lqcd{\Lambda_{\rm QCD}}

\def\GeV{\mbox{GeV}}

\def\mbups{\ensuremath{m_b^{1S}}}

\def\eqn#1{Eq.\ (\ref{#1})}

\def\q{\hat q}
\def\Dlhat{\hat\Delta}
\def\Dfull{{\cal D}}
\def\Dslash{D\hskip -0.65 em /}

\def\gev{{\rm GeV}}

\def\ltapp{\lesssim}
\def\fig#1{Fig.\ \ref{#1}}

\def\xslash#1{{\rlap{$#1$}/}}
\def\dsl{\,\raise.15ex\hbox{/}\mkern-13.5mu D}
\def\half{{\textstyle{1\over 2}}}

\def\ndotq{n \cdot \hat q}
\def\nbardotq{\bar{n} \cdot \hat q}
\def\ndotk{n \cdot \hat k}

\def\ndotl{n \cdot \hat\ell}
\def\shhat{\hat s_H}
\def\btou{\bar{B} \to X_u \ell \bar{\nu}_\ell}
\def\babar{\mbox{\slshape B\kern-0.1em{\smaller A}\kern-0.1em
    B\kern-0.1em{\smaller A\kern-0.2em R}}}
\newcommand{\nn}{\nonumber}
\newcommand{\beq}{\begin{equation}}
\newcommand{\eeq}{\end{equation}}
\newcommand{\beqa}{\begin{eqnarray}}
\newcommand{\eeqa}{\end{eqnarray}}

\begin{document}
\ifpdf
\DeclareGraphicsExtensions{.pdf, .jpg}
\else
\DeclareGraphicsExtensions{.eps, .jpg}
\fi

\preprint{ \hbox{hep-ph/0312366} }

\title{\boldmath Subleading shape function contributions to the hadronic
invariant mass
spectrum in $\bar{B} \to X_u \ell \bar{\nu}_\ell$ decay\vspace*{8pt}}

\vspace*{1.5cm}

\author{Craig N. Burrell, Michael E. Luke and Alexander R. Williamson}

\affiliation{Department of Physics, University of Toronto,
	60 St.~George Street, Toronto, Ontario,
	Canada M5S 1A7\vspace{4pt}}

\begin{abstract} \vspace*{8pt}

We study the $O(\lqcd/m_b)$ corrections to the singly and doubly differential
hadronic invariant mass spectra $d\Gamma/ds_H$ and $d\Gamma/ds_H dq^2$ in 
$\btou$ decays, and discuss the implications for the extraction
of the CKM matrix element $|V_{ub}|$.   Using simple models for the subleading
shape functions, the effects of subleading operators are estimated to be at
the
few percent level for experimentally relevant cuts.  The subleading
corrections
proportional to the leading shape function are larger, but largely cancel in
the relation between the hadronic invariant mass spectrum and the photon
spectrum in $\bar B\to X_s\gamma$.  We also discuss the applicability of the
usual prescription of convoluting the partonic level rate with the leading
light-cone wavefunction of the $b$ quark to subleading order.

\end{abstract}

\maketitle

\section{Introduction}

The CKM parameter $|V_{ub}|$ is of phenomenological interest both
because it is a basic parameter of the Standard Model and because
of the role it plays in precision studies of $CP$ violation in the $B$ meson
system.   Currently, the theoretically cleanest determinations of $|V_{ub}|$
come from inclusive semileptonic decays, which are not sensitive to the
details of hadronization.

For sufficiently inclusive observables, inclusive decay rates may be written
as
an expansion
in local operators \cite{operefs}.  The leading order result corresponds to
the
decay of a free $b$ quark
to quarks and gluons, while the subleading corrections, proportional to powers
of $\lqcd/m_b$,
describe the deviations from the parton model.   Up to $O(\lqcd^2/m_b^2)$,
only
two operators arise,
\begin{equation} \label{lambdas}
\lambda_1\equiv {1\over 2m_B}\langle \bar{B} |\bar{h}_v (iD)^2 h_v| \bar{B}
\rangle,\
\lambda_2(\mu)\equiv {1\over 6m_B}\langle \bar{B} |\bar{h}_v \sigma^{\mu\nu}
G_{\mu\nu}
 h_v| \bar{B} \rangle.
\end{equation}
The $B-B^*$ mass splitting determines $\lambda_2(m_b)\simeq 0.12\,\GeV^2$,
while a recent fit to moments of the charged lepton spectrum in
semileptonic $b\to c$ decay obtained \cite{cleolepton}
\begin{equation}
\mbups=4.82\pm 0.07_E\pm 0.11_T\,\GeV,\ \ \lambda_1=-0.25\pm 0.02_{ST}\pm
0.05_{SY}\pm 0.14_T\,\GeV^2
\end{equation}
where $\mbups$ is the short-distance ``1S mass" of the $b$ quark
\cite{upsexpansion1,upsexpansion2}.
(Moments of other spectra give similar results \cite{moments1,moments2}.)
These uncertainties correspond to an uncertainty of $\sim 5\%$ in the relation
between $|V_{ub}|$ and the inclusive $\btou$ width
\cite{upsexpansion1,burels}.

Unfortunately, the semileptonic $b\to u$ decay rate is difficult to measure
experimentally, because of the large background from charmed final states.
As a result, there has been much theoretical and experimental interest in the
decay rate in restricted regions of phase space where the charm
background is absent.  Of particular interest have been the large lepton
energy
region,  $E_\ell > (m_B^2-m_D^2)/2m_B$, the low
hadronic invariant mass region, $m_X \equiv \sqrt{s_H}< m_D$ \cite{lowmX}, the
large lepton invariant mass region $q^2>(m_B-m_D)^2$ \cite{BLLqsq}, and
combinations of these \cite{BLLmixed}.  The charged lepton cut is the easiest
to implement experimentally, while the hadronic mass cut has the
advantage that it contains roughly $80\%$ of the semileptonic rate \cite{dFN}.
However, in both cases the kinematic cuts constrain the final hadronic
state to consist of energetic, low-invariant mass hadrons, and the local OPE
breaks down (this is not the case for the large $q^2$ region or for
appropriately chosen mixed cuts).  In this case, the relevant spectrum is
determined at leading order in $\lqcd/m_b$ by the light-cone distribution
function of the $b$ quark in the meson \cite{shapeleading1,shapeleading2},
\begin{equation}
f(\omega)\equiv \frac{\langle \bar{B}|\bar b\;\delta(\omega+i n \cdot \hat
D)\;b| \bar{B} \rangle}{2 m_B}
\end{equation}
where $n^\mu$ is a light-like vector, and hatted variables are normalized to
$m_b$: $\hat D^\mu\equiv D^\mu/m_b$.\footnote{Because in our definition of
$f(\omega)$
its argument is dimensionless, $f(\omega)$ differs 
by a factor of $m_b$ from the usual definitions in the literature.}
$f(\omega)$ is often referred to as the shape function, and corresponds to
resumming
an infinite series of local operators in the usual OPE.  
The physical spectra are determined by
convoluting the shape function with the appropriate kinematic functions:
\begin{eqnarray}
\label{eLspect}{1\over\Gamma_0}{d\Gamma\over d\hat E_\ell}(\btou)
	&=&4\int \theta(1-2 \hat E_\ell-\omega)
	f(\omega)\;d\omega+\dots\\
\label{sHspect}{1\over\Gamma_0}{d\Gamma\over d\shhat}(\btou)&=&\int
{2\shhat^2(3\omega-2\shhat)\over\omega^4}\theta(\omega-\shhat)
f(\omega-\hat\Delta)\;d\omega+\dots
\end{eqnarray}
where $1-2 \hat E_\ell \ltapp \lqcd/m_b$, $\shhat\ltapp\lqcd/m_b$ and
$\Delta\equiv m_B-m_b$.

Since $f(\omega)$ also determines the shape of the photon spectrum in $\bar{B}
\to
X_s\gamma$ at leading order,
\begin{equation} \label{bsgLeading}
{1\over\Gamma_0^s}{d\Gamma\over d \hat E_\gamma}(\bar{B}\to X_s\gamma)=2 f(1-2
\hat
E_\gamma)+\dots \label{eGspect}
\end{equation}
there has been much interest in extracting $f(\omega)$ from radiative $B$
decay
and applying it to semileptonic decay.
However, the relations (\ref{eLspect}--\ref{eGspect}) hold only at tree level
and at leading order in $\lqcd/m_b$, so a precision determination of
$|V_{ub}|$ requires an understanding of the size of the corrections.
Radiative corrections were considered in
\cite{shapeleading1,shapeleading2,shaperadiative,Bauer:2003pi},
while $O(\lqcd/m_b)$ corrections have been studied more
recently in \cite{blm01, blm02, mn02,llw02}.  In \cite{blm01}, the nonlocal
distribution functions arising at subleading order were enumerated, and their
contribution to $\bar{B}\to X_s\gamma$ decay was studied.  In \cite{blm02},
the corresponding corrections to the lepton endpoint spectrum in $\btou$
decay were studied, and it was shown that these effects were
potentially large.  Similar results were
obtained in \cite{llw02}, where the
sub-subleading contribution from annihilation graphs was also shown to be
large.  In this paper, we study the subleading corrections to the hadronic
invariant mass spectrum in semileptonic $b\to u$ decay, and estimate the
theoretical uncertainties introduced by these terms.   In addition, we present
results
for the doubly differential spectrum $d\Gamma/ds_H d q^2$ at leading and
subleading
order.  

\section{Matching Calculation}

\subsection{The full theory spectrum}

In the shape function region the final hadronic
state has large energy but small invariant mass, and so its momentum lies
close
to the light-cone.  It is therefore convenient to introduce two light-like
vectors
$n^\mu$ and $\bar{n}^\mu$ related to the velocity of the heavy meson $v^\mu$
by $v^\mu = \half(n^\mu+\bar{n}^\mu)$, and satisfying
\begin{equation}
n^2 = \bar{n}^2 = 0, \quad v\cdot n = v\cdot \bar{n} = 1, \quad n\cdot
\bar{n} = 2.
\end{equation}
In the frame in which the $B$ meson is at rest,
these vectors are given by $n^\mu=(1,0,0,1)$, $\bar
n^\mu=(1,0,0,-1)$ and $v^\mu=(1,0,0,0)$.
The projection of an arbitrary four-vector $a^\alpha$ onto the directions
which
are perpendicular to the
light-cone is given by $ a_\perp^\alpha = g^{\alpha\beta}_\perp a_\beta$,
where
\begin{equation}
g^{\mu \nu}_\perp = g^{\mu \nu}-\frac{1}{2} \left(n^\mu \bar{n}^\nu +
\bar{n}^\mu n^\nu  \right)\,.
\end{equation}
Choosing our axes such that the momentum transfer to the leptons $\vec q$ is
in
the $-\vec n$ direction, we can write
$q^\mu = \half n\cdot q\; \bar{n}^\mu + \half \bar n \cdot q\; n^\mu $, the
decay rate takes a particularly simple form in terms of the variables $n\cdot
q$
and
$\bar n \cdot q$:
\begin{equation} \label{dGndotq}
d\Gamma(\bar{B} \to X_u \ell \bar{\nu}_\ell) = 96 \pi\; \Gamma_0\; W_{\mu\nu}
	L^{\mu\nu}
	(\ndotq-\nbardotq)^2   \theta(\nbardotq) \theta(\ndotq - \nbardotq)
	d\ndotq\; d\nbardotq
\end{equation}
where
\begin{equation} \label{partonicrate}
\Gamma_0 = \frac{G_F^2 |V_{ub}|^2 m_b^5}{192 \pi^3}.
\end{equation}
The hadron tensor $W^{\mu\nu}$ is defined by
\begin{eqnarray} \label{hadrontensor}
W^{\mu\nu} \equiv -\frac{1}{\pi} \mbox{Im}\left(-i \int d^4 x\, e^{-i q\cdot
x}
	\frac{\langle \bar{B}|T[J_L^{\dagger \mu}(x) J_L^\nu(0)] |\bar{B}\rangle}{2
m_B} \right),
	\end{eqnarray}
where the weak current is $J_L^\mu=\bar u\gamma^\mu(1-\gamma_5)b$,
while the lepton tensor is
\begin{equation}
	L^{\mu\nu} \equiv \int \; d\Pi_2(q; p_\ell, p_\nu )  \mbox{Tr}[
\xslash{p}_\nu
\gamma^\mu \xslash{p}_\ell \gamma^\nu P_L ]  = \frac{1}{12 \pi} (
q^\mu q^\nu - q^2
g^{\mu\nu} )
\end{equation}
and $P_L\equiv \half(1-\gamma_5)$.

To calculate the hadronic invariant mass spectrum we switch to the variables
$(s_H, q^2)$.  These are related to the variables in \eqn{dGndotq} by
\begin{eqnarray} \label{sHdefinition}
s_H &=& (m_B - n\cdot q)(m_B - n\cdot\bar q) \nn \\
	&=& (m_b + \Delta - n\cdot q) (m_b+\Delta - \bar n \cdot q) \\
q^2 &=& n\cdot q \; \bar n \cdot q
\end{eqnarray}
and
\begin{equation}
\frac{d\Gamma}{dn\cdot q\, d\bar n \cdot q}= \sqrt{((m_b+\Delta)^2 + q^2
-s_H)^2 - 4 (m_b+\Delta)^2 q^2} \frac{d\Gamma}{ds_H \, dq^2}.
\end{equation}
Here $\Delta = m_B - m_b$ is the difference between the $B$ meson mass
and the $b$ quark mass.  It is ${\cal O}(\lqcd)$ and has an expansion in terms
of HQET parameters
\begin{equation}\label{deltaexpansion}
\Delta = \bar{\Lambda} - \frac{\lambda_1 + 3 \lambda_2}{2 m_b} + \cdots\ .
\end{equation}
Since $\Delta$ simply enters in the definition of $s_H$, it is unrelated to the
$1/m$ expansion
in the OPE, so we will not expand it via \eqn{deltaexpansion}.
With this change of variables, we define the correlator $T(s_H, q^2)$ by
\begin{equation}
\frac{1}{\Gamma_0} \frac{d\Gamma}{ds_H\;dq^2} ={1\over 2 m_B}\langle \bar{B}
|T(s_H,
q^2)| \bar{B} \rangle.
\end{equation}

In Ref.\ \cite{blm01} a nonlocal expansion was performed for the hadron tensor
$W^{\mu\nu}$, based on the power counting
\begin{eqnarray} \label{qscaling}
(m_b v - q)\cdot \bar n &=& m_b - q\cdot \bar n \sim  {\cal O}(m_b), \nonumber
\\
(m_b v - q)\cdot n &=& m_b - q\cdot n \sim {\cal O}(\lqcd), \\
k^\mu&\sim&{\cal O}(\lqcd).\nonumber
\end{eqnarray}
 where the heavy quark momentum is defined as $p_b^\mu=m_b v^\mu+k^\mu$.
However, the limits of phase space integration in \eqn{dGndotq}
include regions of phase space where this power counting is violated.  Hence,
to keep our power counting consistent, we do not perform a nonlocal
OPE for $W^{\mu\nu}$, but rather for $T(s_H, q^2)$.  In these variables, the
shape function region corresponds to the region of low invariant mass,
\begin{equation} \label{scaling}
s_H \sim {\cal O}(\lqcd m_b).
\end{equation}
Since $\Delta\sim\lqcd$ and $k^\mu\sim \lqcd$, expanding the light quark
propagator in powers of $\lqcd/m_b$ gives at leading order
\begin{equation}
\frac{i \xslash{p}_u}{p_u^2} = \frac{i \xslash{n}}{2 m_b}
\frac{(1-\q^2)}{[\shhat - (\Dlhat-\ndotk)(1-\q^2)]} + \cdots\
\end{equation}
(where $\Dlhat\equiv\bar\Lambda/m_b$).
Since both terms in the denominator are $O(\lqcd/m_b)$, $T(s_H, q^2)$ cannot
be
expanded in powers of $k^\mu$ and matched onto local operators
(unless we also are restricted to large $q^2$, such that $1-\q^2\ll 1$, in
which case the second term in the denominator is subleading, and a local OPE
may be performed
\cite{BLLqsq, BLLmixed}).  Instead, the OPE takes the schematic form
\begin{equation}\label{nlOPE}
T(\shhat,\q^2)=\sum_n \int  C_n(\omega,\shhat,\q^2) {\cal O}_n(\omega) d\omega
\end{equation}
where the ${\cal O}_n(\omega)$'s are bilocal operators in which the two points
are separated along the light cone.

\subsection{Nonlocal operators}

In Refs.\ \cite{blm01,blm02}, it was shown that up to subleading order in
$\lqcd/m_b$, the following operators were required in the OPE (\ref{nlOPE}):
\begin{eqnarray} \label{OandPoperators}
O_0(\omega) &=& \bar{h}_v \delta(\omega+i n \cdot \hat{D}) h_v \\
O_1^\mu(\omega) &=& \bar{h}_v \{ i \hat{D}^\mu, \delta(\omega+i n \cdot
\hat{D}) \}
h_v \nn \\
O_2^\mu(\omega) &=& \bar{h}_v [ i \hat{D}^\mu, \delta(\omega+i n \cdot \hat{D})
]
h_v \nn \\
O_3(\omega) &=& \int \!d \omega_1 d \omega_2 \, \delta(\omega_1, \omega_2;
\omega)
\bar{h}_v \delta(in \cdot \hat{D} \!+\! \omega_2) g^{\mu\nu}_\perp \{ i
\hat{D}_\mu, i \hat{D}_\nu \}
\delta(i n \cdot \hat{D} \!+\! \omega_1)  h_v, \nn\\
O_4(\omega) &=& - \int \!d \omega_1d \omega_2\,\delta(\omega_1, \omega_2;
\omega)
\bar{h}_v \delta(in \cdot \hat{D} \!+\! \omega_2)\epsilon_\perp^{\mu\nu} [ i
\hat{D}_\mu, i \hat{D}_\nu ] \delta(in \cdot \hat{D} \!+\!
\omega_1)  h_v \,,\nn \\
P_0^\eta(\omega) &=& \bar{h}_v \delta(\omega+i n \cdot \hat{D}) \gamma^\eta
\gamma_5 h_v \nn \\
P_1^{\mu\eta}(\omega) &=& \bar{h}_v \{ i \hat{D}^\mu, \delta(\omega+i n \cdot
\hat{D}) \} \gamma^\eta \gamma_5 h_v \nn \\
P_2^{\mu\eta}(\omega) &=& \bar{h}_v [ i \hat{D}^\mu, \delta(\omega+i n \cdot
\hat{D}) ] \gamma^\eta \gamma_5 h_v \nn \\
P_3^\eta(\omega) &=& \int \!d \omega_1 d \omega_2 \, \delta(\omega_1, \omega_2;
\omega)
\bar{h}_v \delta(in \cdot \hat{D} \!+\! \omega_2) g^{\mu\nu}_\perp \{ i
\hat{D}_\mu, i \hat{D}_\nu \} \delta(i n \cdot \hat{D} \!+\! \omega_1)
\gamma^\eta \gamma_5 h_v, \nn\\
P_4^\eta(\omega) &=& - \int \!d \omega_1d \omega_2\,\delta(\omega_1, \omega_2;
\omega)
\bar{h}_v \delta(in \cdot \hat{D} \!+\! \omega_2)\epsilon_\perp^{\mu\nu} [ i
\hat{D}_\mu, i \hat{D}_\nu ] \delta(in \cdot \hat{D} \!+\!
\omega_1) \gamma^\eta \gamma_5  h_v \,,\nn
\end{eqnarray}
where the $h_v$'s are heavy quark fields in HQET, and we have defined
\begin{equation}
\delta(\omega_1, \omega_2; \omega) = \frac{\delta(\omega-\omega_1)-
\delta(\omega-\omega_2)}{\omega_1 - \omega_2}
\end{equation}
and
\begin{equation}
\epsilon_\perp^{\alpha\beta} = \epsilon^{\alpha \beta \sigma \rho} v_\sigma
n_\rho.
\end{equation}
These definitions differ slightly from the definitions in Refs.\
\cite{blm01,blm02},
because we have chosen to normalize all momenta to $m_b$, to keep the resulting
formulas
simpler.

\begin{figure}[tbp]
\centerline{\includegraphics[width=6in]{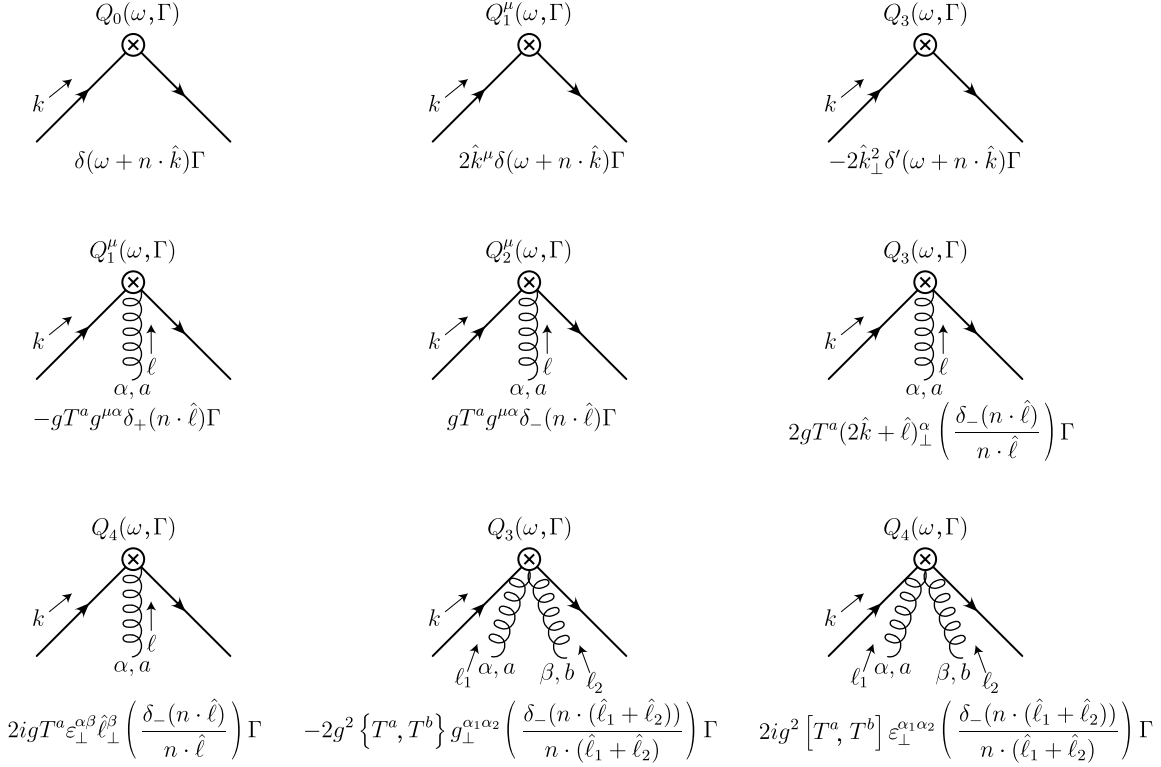}}
\caption{Feynman rules for the operators $Q_i(\omega, \Gamma)$ in $n\cdot A=0$
gauge.  We have defined $\delta_\pm(x) =
\delta(\omega+\ndotk+x) \pm \delta(\omega+\ndotk)$.}
\label{qrules}
\end{figure}

It is convenient to calculate the matching conditions onto a slightly
different set of operators, defined in terms of full QCD $b$ quark fields:
\begin{eqnarray} \label{subleadingoperators}
Q_0(\omega,\Gamma) &=& \bar{b} \delta(i n \cdot \hat{\Dfull} + \omega)
	\Gamma b\, \\\
Q_1^{\mu}(\omega,\Gamma)&=&
\bar{b} \left\{i \hat{\Dfull}^\mu,\delta(in \cdot \hat{\Dfull} +
\omega)\right\}
	\Gamma b,\nonumber\\
Q_2^{\mu}(\omega,\Gamma) &=&
\bar{b} \left[i \hat{\Dfull}^\mu,\delta(in \cdot \hat{\Dfull} + \omega)\right]
	\Gamma b ,\nn \\
Q_3(\omega,\Gamma) &=&
\!\!\int \!d \omega_1 d \omega_2 \, \delta(\omega_1, \omega_2; \omega)
\bar{b} \delta(i n \cdot \hat{\Dfull} \!+\! \omega_2) g^{\mu\nu}_\perp \{ i
\hat{\Dfull}_\mu, i \hat{\Dfull}_\nu \} \delta(i n \cdot \hat{\Dfull} \!+\!
\omega_1)
\Gamma b, \nn\\
Q_4(\omega,\Gamma) &=&
-\!\!\int \!d \omega_1d \omega_2\,\delta(\omega_1, \omega_2; \omega)
\bar{b} \delta(in \cdot \hat{\Dfull} \!+\! \omega_2)\epsilon_\perp^{\mu\nu} [ i
\hat{\Dfull}_\mu, i \hat{\Dfull}_\nu ] \delta(in \cdot \hat{\Dfull} \!+\!
\omega_1) \Gamma b \,.\nn
\end{eqnarray}
We have defined
\begin{equation}
i\Dfull^\mu\equiv iD^\mu-m_b v^\mu
\end{equation}
so that $i\Dfull^\mu$ acting on the $b$ fields just bring down factors of the
residual momentum $k^\mu$.
The Feynman rules for the $O_i$'s and $P_i$'s are given in \cite{blm01,
blm02}.
The rules  for the $Q_i$'s are given in $n\cdot A=0$ gauge in
\fig{qrules}, where we have defined
\begin{equation} \label{deltapm}
\delta_\pm(x) = \delta(\omega+\ndotk+x) \pm \delta(\omega+\ndotk).
\end{equation}
%
%
It is simpler to match onto the $Q_i$'s initially since this matching does not
require us to relate the QCD quark fields to HQET quark fields.  However,
because the additional symmetries of HQET reduce
the number of independent functions needed
to parametrize the matrix elements, it is convenient to then express the
$Q_i$'s in
terms of the $O_i$'s and $P_i$'s.  For an
arbitrary Dirac structure $\Gamma$ we have
\begin{eqnarray}\label{quarkexpand}
\bar{b} \delta(\omega + i n \cdot \hat{\Dfull}) \Gamma b &=&
	\bar{h}_v \left(1 + \frac{i \Dslash}{2 m_b}\right)
	\delta(\omega + i n \cdot \hat{D}) \Gamma
	\left(1 + \frac{i \Dslash}{2 m_b}\right) h_v + \cdots\\
&=& \frac{1}{2} \mbox{Tr}[\Gamma P_+] O_0(\omega)
	-\frac{1}{2}\mbox{Tr}[\Gamma s_\eta] P_0^\eta(\omega) \nn \\
	&+&\frac{1}{8} \left( \mbox{Tr}[\gamma_\lambda \Gamma
P_+]+\mbox{Tr}[\Gamma \gamma_\lambda P_+]
\right) O_1^\lambda(\omega) \nn \\
	&+&\frac{1}{8} \left( \mbox{Tr}[\gamma_\lambda \Gamma
P_+]-\mbox{Tr}[\Gamma \gamma_\lambda P_+]
\right) O_2^\lambda(\omega) \nn \\
	&-&\frac{1}{8} \left( \mbox{Tr}[\gamma_\lambda \Gamma
s_\eta]+\mbox{Tr}[\Gamma \gamma_\lambda s_\eta]
\right) P_1^{\lambda \eta}(\omega) \nn \\
	&-&\frac{1}{8} \left( \mbox{Tr}[\gamma_\lambda \Gamma
s_\eta]-\mbox{Tr}[\Gamma \gamma_\lambda s_\eta]
\right) P_2^{\lambda \eta}(\omega) + \cdots \nn
\end{eqnarray}
where
\begin{equation}
P_+ = \half (1 + \xslash{v})\hspace{1cm}\mbox{and}\hspace{1cm}s^\eta = P_+
\gamma^\eta \gamma_5 P_+
\end{equation}
and we have used the fact that
\begin{equation}
\bar{h}_v \Gamma h_v = \frac{1}{2} \mbox{Tr}[\Gamma P_+]
\bar{h}_v h_v - \frac{1}{2} \mbox{Tr}[\Gamma s_\eta]
\bar{h}_v \gamma^\eta \gamma_5 h_v.
\end{equation}
For our purposes, we will only need the case $\Gamma =
\gamma_\sigma P_L$, which allows us to write
\begin{eqnarray} \label{Q0toHQET}
Q_0(\omega,\gamma_\sigma P_L) &=&
\frac{1}{2} v_\sigma O_0(\omega) - \frac{1}{2} ( g_{\sigma \eta}-v_\sigma
v_\eta ) P_0^\eta(\omega) \\
&+& \frac{1}{4} g_{\lambda\sigma} O_1^\lambda(\omega)
 -  \frac{1}{4} ( g_{\sigma\eta} v_\lambda - g_{\lambda\eta} v_\sigma )
P_1^{\lambda\eta}(\omega) \nn \\
&+& \frac{1}{4} i \epsilon_{\sigma\lambda\eta\rho} v^\rho
P_2^{\lambda\eta}(\omega)
 + \cdots\nn
\end{eqnarray}
where the first line gives the leading order relation and subsequent lines
contain
the subleading correction.

Similar relations may be derived for the subleading operators,
though in these cases it is not necessary to consider the
subleading terms in the relation between the QCD operator and
the HQET operator, such terms being of higher order overall.
Thus we have
\begin{eqnarray} \label{QitoHQET}
Q_1^{\mu}(\omega, \gamma_\sigma P_L)
	&=& \half v_\sigma O_1^\mu(\omega)-
	\half (g_{\eta\sigma}-v_\eta v_\sigma) P_1^{\mu\eta}(\omega) + \cdots \nn \\
Q_2^{\mu}(\omega, \gamma_\sigma P_L)
	&=& \half v_\sigma O_2^\mu(\omega)-
	\half (g_{\eta\sigma}-v_\eta v_\sigma) P_2^{\mu\eta}(\omega) + \cdots \\
Q_3(\omega, \gamma_\sigma P_L)
	&=& \half v_\sigma O_3(\omega)-
	\half (g_{\eta\sigma}-v_\eta v_\sigma) P_3^{\eta}(\omega) + \cdots \nn \\
Q_4(\omega, \gamma_\sigma P_L)
&=& \half v_\sigma O_4(\omega)-
	\half (g_{\eta\sigma}-v_\eta v_\sigma) P_4^{\eta}(\omega)+ \cdots\ . \nn
\end{eqnarray}

The leading and subleading operators can then be
completely parametrized in terms of five functions \cite{blm01}:
\begin{eqnarray}\label{Omats}
\langle \bar{B} | O_0(\omega) | \bar{B} \rangle &=& 2 m_B \left( f(\omega) +
\frac{t(\omega)}{2}\right)  \nn \\
\langle \bar{B} | O_1^\mu(\omega) | \bar{B} \rangle &=& 2 m_B (n-\bar{n})^\mu
\omega
f(\omega) \nn \\
\langle \bar{B} | O_3(\omega) | \bar{B} \rangle &=& 4 m_B G_2(\omega) \\
\langle \bar{B} | P_2^{\mu\eta}(\omega) | \bar{B} \rangle &=& -2 m_B
i \epsilon_\perp^{\mu\eta}  h_1(\omega) \nn \\
\langle \bar{B} | P_4^{\eta}(\omega) | \bar{B} \rangle &=& 4 m_B (v-n)^\eta
H_2(\omega) \nn
\end{eqnarray}
(once again, unlike in \cite{blm01}, these are defined here in terms of
dimensionless arguments).
The matrix elements of the other operators vanish.


\subsection{Matching Conditions}
\label{matching_calculation}

\begin{figure}[tbp]
\centerline{\includegraphics[width=3.5in]{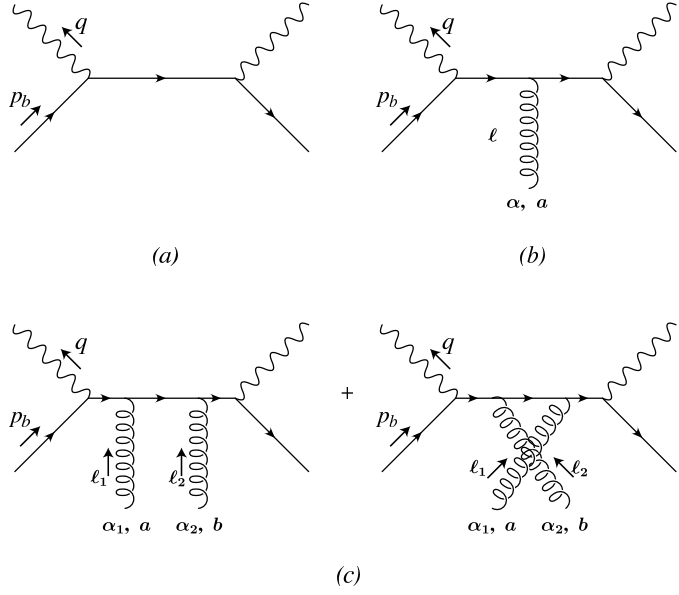}}
\caption{Full-theory forward scattering diagrams.}
\label{fig:FullForwardScatteringDiagrams}
\end{figure}

The Wilson coefficients $C_i(\omega)$ of
the operators in (\ref{nlOPE}) are obtained by taking partonic matrix
elements of both sides of the OPE.  In particular we take zero-,
one-, and two-gluon matrix elements, which corresponds
to calculating the imaginary parts of the full-theory
forward-scattering diagrams in
Figure \ref{fig:FullForwardScatteringDiagrams}, multiplying by the lepton
tensor $L^{\mu\nu}$ and appropriate phase space factors and matching them
onto linear combinations of the
effective diagrams.  (The matching conditions may be completely determined
from
just the zero-gluon and one-gluon
matrix elements, but we have calculated the rest as a check of the results.)

The lepton tensor has the expansion
\begin{eqnarray} \label{leptontensor}
	L^{\mu\nu} &=&  \frac{1}{12 \pi} ( q^\mu q^\nu - q^2
g^{\mu\nu} ) \nn \\
	&=& \frac{m_b^2}{48\pi} \left\{ \bar{n}^\mu \bar{n}^\nu
	+ \q^2 (n^\mu \bar{n}^\nu + \bar{n}^\mu n^\nu) + \q^4 n^\mu n^\nu
	- 4 \q^2 \; g^{\mu\nu} \right\} \\
	&\,&\hspace{0.3cm} -\frac{m_b^2}{24\pi} \frac{(\shhat - \Dlhat
(1-\q^2))}{(1-\q^2)}
	\left\{ \bar{n}^\mu \bar{n}^\nu - \q^4 n^\mu n^\nu \right\} + \cdots \nn.
\end{eqnarray}
(where we have used the decomposition $\q^\mu = \ndotq \bar{n}^\mu/2+
\nbardotq
n^\mu/2$), while the phase space factors give
\begin{eqnarray}\label{phasespace}
  &&{(\ndotq-\nbardotq)^2\over\sqrt{((1+\Dlhat)^2 + \q^2 -\shhat)^2 - 4
(1+\Dlhat)^2 \q^2}} \theta(\ndotq-\nbardotq)\theta(\nbardotq)\nonumber\\
  &&\qquad=(1-\q^2)\theta(\q^2)\theta(1-\q^2)\nonumber\\
  &&\qquad-{(1+\q^2)\shhat-2\q^2(1-\q^2)\Dlhat \over
1-\q^2}\theta(\q^2)\theta(1-\q^2)-2\shhat\delta(1-\q^2)+\dots
 \end{eqnarray}

The zero-gluon diagram in Figure \ref{fig:FullForwardScatteringDiagrams}(a) 
gives the amplitude
\begin{equation}
i A_0 = i \gamma^\mu \frac{\xslash{p}_u}{p_u^2} \gamma^\nu P_L.
\end{equation}
Taking the imaginary part of this amplitude gives
\begin{eqnarray} \label{A0}
-\frac{m_b}{\pi} \mbox{Im}[A_0] &=& \gamma^\mu\xslash{\hat{p}}_u\gamma^\nu P_L
\delta(\hat{p}_u^2) \nn \\
	&=& \frac{1}{2}\gamma^\mu\xslash{n}\gamma^\nu P_L \left[ (1-\q^2)
\delta(h(\ndotk)) \right. \\
	&\,& \hspace{0.2cm} + \left. \left\{ \frac{\shhat}{1-\q^2}
	(\Dlhat (1-\q^4) - \shhat \q^2) - (1-\q^2) \hat{k}_\perp^2 \right\}
\delta'(h(\ndotk)) \right] \nn \\
	&\,& \hspace{0.2cm} + \gamma^\mu\xslash{\hat{k}}_\perp\gamma^\nu P_L
\delta(h(\ndotk)) + \cdots \nn
\end{eqnarray}
where we have expanded the amplitude to subleading order using
(\ref{scaling}) and we have simplified the expression by integrating
by parts.
The function $h(x)$ appearing in (\ref{A0}) is
\begin{equation} \label{h}
h(x) = \shhat - (\Dlhat - x)(1-\q^2).
\end{equation}
Multiplying this result by the lepton tensor (\ref{leptontensor}) and phase
space factors (\ref{phasespace}),
and expanding to subleading order we find
\begin{equation}
\langle b| T(\shhat, \q^2) | b\rangle= \sum_n \int d\omega\;
	\tilde C^\sigma_n(\omega,\shhat, \q^2) \langle b | Q_n(\omega,\gamma^\sigma
P_L)| b\rangle
\end{equation}
where
\begin{eqnarray}\label{wilson1}
\tilde C_0^\sigma(\omega,\shhat, \q^2) &=& 4(1-\q^2)^2 (2 \q^2
n^\sigma+\bar{n}^\sigma)
	 \theta(\q^2) \theta(1-\q^2) \delta(h(-\omega))  \nn \\
	&\,&\hspace{0.5cm} + 8 (1-\q^2) ((\Dlhat(1-3\q^2)- \omega \q^2 (3-\q^2))
n^\sigma \nn \\
	&\,&\hspace{1cm}
	- (\Dlhat + \omega( 2 - \q^2 )) \bar{n}^\sigma ) \theta(\q^2) \theta(1-\q^2)
\delta(h(-\omega))  \nn \\
	&\,&\hspace{0.5cm} + 4 \Dlhat \bar{n}^\sigma \delta(\q^2) \delta(h(-\omega))
+
\cdots \\
\tilde C_1^{\mu\sigma}(\omega,\shhat, \q^2) &=& 4 \q^2 (1-\q^2)
g_\perp^{\mu\sigma}
	\theta(\q^2) \theta(1-\q^2) \delta(h(-\omega)) + \cdots \nn \\
\tilde C_3^{\sigma}(\omega,\shhat, \q^2) &=& -2 (1-\q^2) (2 \q^2
n^\sigma+\bar{n}^\sigma)
	\theta(\q^2) \theta(1-\q^2) \delta(h(-\omega)) + \cdots \nn.
\end{eqnarray}

In order to determine the other matching coefficients, we
calculate the one-gluon amplitude in Figure
\ref{fig:FullForwardScatteringDiagrams}(b).  Defining $\ell$ to be
the incoming gluon momentum, we have
\begin{equation}
i A_1 = i g T^a
	\frac{\gamma^\mu(\xslash{p}_u + \xslash{\ell}) \gamma^\alpha
\xslash{p}_u\gamma^\nu P_L}{(p_u+\ell)^2 p_u^2}.
\end{equation}
where $(\alpha,a)$ are, respectively, the Lorentz and
colour indices of the gluon field.

Taking into account the two cuts which result from taking
Im$[A_1]$ and scaling the gluon momentum as $\ell^\alpha \sim {\cal
O}(\lqcd)$,
we obtain, after expanding to leading order in $n\cdot A=0$ gauge,
\begin{eqnarray} \label{A1}
-\frac{m_b^2}{\pi} \mbox{Im}[A_1] &=& -\frac{g T^a}{4}
	\gamma^\mu\left\{ 2 \gamma^\alpha_\perp \tilde{\delta}_+(\ndotl)
	+ 2 \xslash{n} (2 \hat{k} + \hat{\ell})_\perp^\alpha
		\left( \frac{\tilde{\delta}_-(\ndotl)}{\ndotl} \right) \right. \\
	&+& \left. 2 i \epsilon_\perp^{\alpha\beta} \gamma_\beta \gamma_5
		\tilde{\delta}_-(\ndotl)
	+ 2 i \epsilon_\perp^{\alpha\beta} \hat{\ell}_{\perp \beta}
		\xslash{n} \gamma_5
		\left( \frac{\tilde{\delta}_-(\ndotl)}{\ndotl} \right) \right\}\gamma^\nu P_L
+ \cdots\nn
\end{eqnarray}
where, in analogy with (\ref{deltapm}), we have defined
\begin{equation}
\tilde{\delta}_\pm(x) = \delta(h(\ndotk + x)) \pm \delta(h(\ndotk)).
\end{equation}
Again, multiplying by the lepton tensor and phase space factors gives
\begin{equation}\label{onegluonT}
\langle b | T(\shhat,\q^2) | b \, g \rangle= \sum_n \int
d\omega\;
	\tilde C^\sigma_n(\omega,\shhat, \q^2) \langle b |
Q_n(\omega,\gamma^\sigma P_L)| b \, g \rangle.
\end{equation}
Part of (\ref{A1}) is reproduced by combining the
Wilson coefficients (\ref{wilson1}) determined earlier with the
one-gluon Feynman rules for $Q_{1,3}(\omega,\gamma^\sigma P_L)$, while the
remainder
corresponds to matrix elements of $Q_{2,4}(\omega,\gamma^\sigma P_L)$ with the
coefficients
\begin{eqnarray} \label{wilson2}
\tilde C_2^{\mu\sigma}(\omega,\shhat,\q^2) &=& 4 \q^2 (1-\q^2)
i \epsilon_\perp^{\mu\sigma}
	\theta(\q^2) \theta(1-\q^2) \delta(h(-\omega)) + \cdots \nn \\
\tilde C_4^{\sigma}(\omega,\shhat,\q^2) &=& -2(1-\q^2) (2 \q^2
n^\sigma+\bar{n}^\sigma)
	\theta(\q^2) \theta(1-\q^2) \delta(h(-\omega)) + \cdots\ .
\end{eqnarray}

The final matrix element to evaluate is the two-gluon
amplitude, \fig{fig:FullForwardScatteringDiagrams}(c).  The amplitude is
\begin{eqnarray}
i A_2 &=& i g^2 \gamma^\mu\left\{ T^a T^b \frac{(\xslash{p}_u +
\xslash{\ell}_1+\xslash{\ell}_2) \gamma^{\alpha_1}
	(\xslash{p}_u + \xslash{\ell}_2) \gamma^{\alpha_2} \xslash{p}_u}
	{(p_u+\ell_1+\ell_2)^2 (p_u+\ell_2)^2 p_u^2} \right.\\
	&\,&\hspace{2cm}+ \left. T^b T^a \frac{(\xslash{p}_u +
\xslash{\ell}_1+\xslash{\ell}_2) \gamma^{\alpha_2}
	(\xslash{p}_u + \xslash{\ell}_1) \gamma^{\alpha_1}
\xslash{p}_u}{(p_u+\ell_1+\ell_2)^2 (p_u+\ell_1)^2 p_u^2}
	\right\} \gamma^\nu P_L \nn
\end{eqnarray}
so that after cutting the diagrams and expanding to leading order, again in
$n\cdot A=0$ gauge,
we obtain
\begin{equation} \label{A2}
-\frac{m_b^3}{\pi} \mbox{Im}[A_2] = \frac{g^2}{2}\gamma^\mu\xslash{n}\gamma^\nu
P_L
	\left\{  g_\perp^{\alpha_1 \alpha_2} \{ T^a, T^b \}
		- i \epsilon_\perp^{\alpha_1 \alpha_2} [ T^a, T^b] \right\}
		\frac{\tilde{\delta}_-(n \cdot (\hat{\ell}_1+\hat{\ell}_2))}{n \cdot
(\hat{\ell}_1+\hat{\ell}_2)}
		+ \cdots
\end{equation}
The two gluon matrix element of $T(\shhat, \q^2)$ agrees with the results of
(\ref{wilson1}) and (\ref{wilson2}) for $C_3$ and $C_4$; hence, no new
operators are required, as expected.

Integrating these expressions over $q^2$ we obtain the OPE for $d\Gamma/d s_H$
\begin{equation}
\frac{1}{\Gamma_0} \frac{d \Gamma}{d\shhat} = \frac{1}{2m_B}\sum_n
\int_{-\infty}^{\infty} d\omega \;
	C_n^{\sigma}(\omega, \shhat)
	\langle \bar{B} | Q_n(\omega, \gamma_\sigma P_L) | \bar{B} \rangle
\end{equation}
where
\begin{eqnarray} \label{WilsonCoefficients1}
C_0^\sigma(\omega, \shhat) &=&  - \frac{4 \shhat^2 (2 (\shhat-\Dlhat-\omega)
n^\sigma - (\Dlhat+\omega) \bar{n}^\sigma
)}{(\Dlhat+\omega)^4}
	 \theta\left(\shhat \right) \theta\left(\Dlhat+\omega-\shhat \right)\nn \\
	 &+& \frac{8 \shhat}{ (\Dlhat+\omega)^4} \left\{  (\shhat^2 \omega + \shhat
(\omega^2+4 \omega \Dlhat + 3 \Dlhat^2)-
2(\Dlhat+\omega)^3) n^\sigma \right. \nn \\
	&\;&\hspace{0.5cm} \left. - (\Dlhat+\omega)(\Dlhat^2+ 2 \omega \Dlhat +
\omega
(\shhat+\omega)) \bar{n}^\sigma \right\}
	\theta\left( \shhat \right) \theta\left(\Dlhat + \omega - \shhat \right) \nn
\\
	&+& 4 \Dlhat \bar{n}^\sigma  \delta \left(\Dlhat + \omega -\shhat \right) +
\cdots \\
C_1^{\mu\sigma}(\omega, \shhat) &=& -\frac{4 \shhat (\shhat-\Dlhat-\omega)
g^{\mu\sigma}_\perp}{ (\Dlhat+\omega)^3}
	\theta\left( \shhat \right) \theta\left( \Dlhat + \omega - \shhat
\right)+\cdots \nn \\
C_2^{\mu\sigma}(\omega, \shhat) &=& -\frac{4 \shhat (\shhat - \Dlhat - \omega)
i \epsilon^{\mu\sigma}_\perp}{(\Dlhat+\omega)^3}
	 \theta\left( \shhat \right) \theta\left(\Dlhat + \omega -\shhat
\right)+\cdots
	   \nn \\
C_3^{\sigma}(\omega, \shhat) &=& \frac{2 \shhat ( 2 (\shhat-\Dlhat-\omega)
n^\sigma -
	(\Dlhat+\omega) \bar{n}^\sigma)}{(\Dlhat+\omega)^3}
	\theta\left( \shhat \right) \theta\left(\Dlhat + \omega -\shhat
\right)+\cdots
\nn \\
C_4^{\sigma}(\omega, \shhat) &=& \frac{2 \shhat  ( 2 (\shhat-\Dlhat-\omega)
n^\sigma - (\Dlhat+\omega) \bar{n}^\sigma)
}{(\Dlhat+\omega)^3}
	\theta\left( \shhat \right) \theta\left(\Dlhat + \omega - \shhat
\right)+\cdots \nn.
\end{eqnarray}

Finally, relating the $Q_i$'s to the $O_i$'s and $P_i$'s via 
(\ref{Q0toHQET}) and (\ref{QitoHQET}) and taking the matrix elements
(\ref{Omats}),
we obtain the expression for the hadronic invariant mass spectrum:
\begin{eqnarray} \label{sHspectrum}
\frac{1}{\Gamma_0} \frac{d \Gamma}{d \shhat} &=&
	\int_{-\infty}^\infty \;d\omega \left\{
	\frac{2 \shhat^2 (3 \omega - 2 \shhat) f(\omega-\Dlhat)}{\omega^4}
	\theta\left(\shhat\right) \theta\left( \omega-\shhat \right) \right. \\
&+& 2 \Dlhat f(\omega-\Dlhat)
	\theta\left( \shhat \right) \delta \left( \omega- \shhat \right) \nn \\
&+& \left[ \frac{2 \shhat (4 \shhat^2 (\omega-\Dlhat)+ \shhat \omega (7 \Dlhat
- \omega)- 6 \omega^3)
	f(\omega-\Dlhat)}{\omega^4}
\right. \nn \\
&+& \frac{ \shhat^2 (3 \omega - 2 \shhat) t(\omega-\Dlhat)}{\omega^4}
- \frac{2 \shhat (3 \omega - 2 \shhat) G_2(\omega-\Dlhat)}{\omega^3} \nn
\\
&+& \frac{2 \shhat (2 \shhat^2 + \omega \shhat - 2 \omega^2)
h_1(\omega-\Dlhat)}{\omega^4} \nn \\
&-& \left. \left. \frac{2 \shhat (2 \shhat - \omega)
H_2(\omega-\Dlhat)}{\omega^3} \right] \theta\left( \shhat \right)
\theta\left(\omega-\shhat \right) \right\} + \cdots \nn
\end{eqnarray}

\eqn{sHspectrum} is the principal result of this paper.  It may be checked for
consistency with the result obtained via the local OPE by expanding the
matrix elements of the operators (\ref{OandPoperators}) such that
$ i n \cdot D \sim {\cal O}(\lqcd)$.
This gives \cite{blm01}
\begin{eqnarray} \label{fexp}
f(\omega) &=& \delta(\omega)
              - \frac{\lambda_1}{6 m_b^2} \delta '' (\omega) -
                \frac{\rho_1}{18 m_b^3} \delta ''' (\omega) + \cdots
\nonumber\\
\omega f(\omega) &=&  \frac{\lambda_1}{3 m_b^2} \delta'(\omega) +
      \frac{\rho_1}{6 m_b^3} \delta '' (\omega) + \cdots  \nonumber\\
h_1(\omega) &=& \frac{\lambda_2}{m_b^2} \delta'(\omega)
               +\frac{\rho_2}{2 m_b^3} \delta''(\omega) + \cdots  \\
G_2(\omega) &=& -\frac{2 \lambda_1}{3 m_b^2} \delta'(\omega) + \cdots \nn \\
H_2(\omega) &=& -\frac{\lambda_2}{m_b^2} \delta'(\omega) + \cdots \nn \\
t(\omega) &=& - \frac{\lambda_1 + 3 \lambda_2}{m_b^2} \delta ' (\omega)
              + \frac{\tau}{2 m_b^3} \delta '' (\omega) + \cdots \,\nonumber
\end{eqnarray}
where each term in the expansion is of the same order in the shape function
region, but the terms indicated by ellipses are higher order in the local OPE.
The $\lambda_{1,2}$ parameters are defined in (\ref{lambdas}) and the
$\rho_{1,2}$ parameters are defined by
\begin{eqnarray} \label{rhos}
\frac{1}{2 m_B} \langle \bar{B} | \bar{h}_v i D_\alpha i D_\mu i D_\beta h_v |
\bar{B}
\rangle &=&
	\frac{1}{3} (g_{\alpha\beta} - v_\alpha v_\beta) v_\mu \rho_1 \\
\frac{1}{2 m_B} \langle \bar{B} | \bar{h}_v i D_\alpha i D_\mu i D_\beta
s_\delta h_v
| \bar{B} \rangle &=&
	\frac{1}{2} i \epsilon_{\nu\alpha\beta\delta} v^\nu v_\mu \rho_2 \nn.
\end{eqnarray}

 When substituted into the spectrum (\ref{sHspectrum}) and integrated over
$\omega$ we obtain to subleading order
\begin{eqnarray} \label{localsfregion}
\frac{1}{\Gamma_0} \frac{d\Gamma_{\mathrm{local}}}{d\shhat} &=& \left\{ \frac{2
\shhat^2
(3 \Dlhat - 2 \shhat ) }{\Dlhat^4} \theta(\Dlhat-\shhat)
\theta(\shhat) \right. \\
&\,& \hspace{1cm} + \hat{\lambda}_1 \left( \frac{4 \shhat^2 (10 \shhat - 9
\Dlhat)}{3 \Dlhat^6} \theta(\Dlhat-\shhat) \theta(\shhat)  + \frac{2}{3
\Dlhat^2} \delta(\shhat-\Dlhat) \right) \nn \\
&\,& \hspace{1cm}  \left. +
\hat{\rho}_1 \frac{20 \shhat^2 (4 \shhat - 3 \Dlhat)}{3 \Dlhat^7}
\theta(\Dlhat-\shhat) \theta(\shhat)  \right\} \nn \\
&+& \frac{12 \shhat (\shhat-\Dlhat)}{\Dlhat^2} \theta(\Dlhat-\shhat)
\theta(\shhat) + 2 \Dlhat \delta(\shhat - \Dlhat) \nn \\
&+& \hat{\lambda}_1 \left( \frac{\shhat (56 \shhat^2 - 129 \Dlhat \shhat + 36
\Dlhat^2)}{3 \Dlhat^5}  \theta(\Dlhat-\shhat) \theta(\shhat) +
\frac{11}{3\Dlhat} \delta(\shhat-\Dlhat) \right) \nn \\
&+& \hat{\lambda}_2 \left( \frac{\shhat(40 \shhat^2 - 9 \Dlhat \shhat - 12
\Dlhat^2)}{\Dlhat^5} \theta(\Dlhat-\shhat) \theta(\shhat) -
\frac{1}{\Dlhat} \delta(\shhat-\Dlhat) \right) \nn \\
&+& \hat{\rho}_1 \left( \frac{4 \shhat (20 \shhat^2 - 33 \Dlhat \shhat + 3
\Dlhat^2)}{3 \Dlhat^6} \theta(\Dlhat-\shhat) \theta(\shhat) +
\frac{6}{\Dlhat^2} \delta(\shhat-\Dlhat) \right) \nn \\
&+& \hat{\rho}_2 \left( \frac{4 \shhat (10 \shhat^2 + 3 \Dlhat \shhat - 3
\Dlhat^2)}{\Dlhat^6} \theta(\Dlhat-\shhat) \theta(\shhat) -
\frac{8}{\Dlhat^2} \delta(\shhat-\Dlhat) \right) + \cdots \nn
\end{eqnarray}
where the terms in curly brackets are the leading order result, and the other
terms are the subleading order correction.

The local OPE spectrum can be obtained from the double-differential spectrum
$d\Gamma/ds_0 d E_0$ presented in \cite{moments1} and \cite{gremmkapustin}.
After changing variables to $(s_H, E_0)$ and expanding in powers of
$\lqcd/m_b$
(treating $s_H$ as order $\lqcd m_b$), performing the $E_0$ integral we
obtain the local OPE for $d\Gamma/d\shhat$, which exactly reproduces the
result
 (\ref{localsfregion}).

\section{Relation to Previous Work}

At leading order in $1/m_b$, the effects of the distribution function
$f(\omega)$ may be simply included by replacing $m_b$ in the tree-level
partonic rate
\begin{equation}\label{mbsmear}
m_b\to m_b^*=m_b(1-\omega)
\end{equation}
and then convoluting the differential rate $d\Gamma$ with the distribution
function $f(\omega)$ \cite{shapeleading1},
\begin{equation}\label{smearing}
d\Gamma=\int  d\Gamma^{\rm parton}\vert_{m_b\to m_b^*} f(\omega)\; d\omega.
\end{equation}
Because of the leading factor of $m_b^5$ in the rate (\ref{partonicrate}), 
this prescription leads
to large subleading corrections if the factor of $m_b^5$ is included in the
replacement (\ref{mbsmear}).

In Ref.\ \cite{dFN} this prescription was applied to the $s_H$ spectrum,
although
the $m_b^5$ term was not included in the replacement.   This is perfectly
consistent
at leading order, but since other subleading effects were introduced
in Ref.\ \cite{dFN} by the replacement (\ref{mbsmear}), it is instructive to
compare our result (\ref{sHspectrum}) with the results of Ref.\ \cite{dFN},
expanded consistently to subleading order in $1/m_b$.  
At leading order, the results are identical:\footnote{In Ref.\ \cite{dFN} the
upper
limit of integration is $\omega=\sqrt{\shhat}$; however, the difference is
higher order.
In addition, the region $\omega\sim\sqrt{\shhat}\gg \shhat$ is outside the
region of support of the shape function, and so is expected to be suppressed.}
\begin{equation}\label{leadingordersH}
{1\over \Gamma_0} {d\Gamma^{(0)}\over d\shhat}=\int_{0}^\infty A(\shhat,
\omega)
f(\omega-\Dlhat)\;d\omega
\end{equation}
where
\begin{equation}
A(\shhat,\omega)= {2 \shhat^2(3\omega-2 \shhat)\over \omega^4}
\theta(\omega-\shhat)
\end{equation}
At subleading order, the relevant terms in \eqn{sHspectrum} may be written as
\begin{equation}\label{subleadingkin1}
{1\over\Gamma_0}{d\Gamma^{(1)}\over d\shhat}=\int_{0}^\infty \delta A(\shhat,
\omega) f(\omega-\Dlhat)\;d\omega+\dots
\end{equation}
where the ellipses denote subleading shape functions, the effects of which
cannot be reproduced by
the prescription (\ref{smearing}).    We will refer to these corrections as
true subleading
corrections, and the terms arising from $\delta A(\shhat,\omega)$ as kinematic
correction.
The function $\delta A(\omega,\shhat)$ is
\begin{eqnarray}\label{subleadingkin2}
\delta A(\omega,\shhat)&=&\frac{2 \shhat ( \shhat (5 \omega + \Dlhat) - 6
\omega^2) }{\omega^3} \theta(\omega-\shhat)+2\Dlhat\delta(\omega-\shhat)
\nonumber\\
&=&{2 \shhat(8\shhat^2(\Dlhat-\omega)+3
\shhat\omega(5\omega-3\Dlhat)-6\omega^3)\over\omega^4}+
2\Dlhat\delta(\omega-\shhat)\nonumber\\
&&+{10 \shhat^2(2 \shhat-3\omega)(\omega-\Dlhat)\over\omega^4}+{2\shhat^2
(2\shhat-\omega)(\omega-\Dlhat)\over\omega^4}
\end{eqnarray}
The second line of \eqn{subleadingkin2} agrees with the expansion of the
results of Ref.\ \cite{dFN} to subleading order.   The first term in the third
line agrees with the
expansion if the $m_b^5$ factor is also included in the convolution.
 Finally, the last term in \eqn{subleadingkin2} arises from the expansion of
the quark fields in terms of HQET fields in the relation (\ref{quarkexpand}).
Thus, we see that to be consistent to subleading order, one must include the
$m_b^5$ term in the replacement (\ref{mbsmear}).  However, like the
subleading shape functions, the subleading
effects arising from the expansion of the quark fields cannot be reproduced by
this procedure.

\begin{figure}[tbp]
\centerline{\includegraphics[width=4.5in]{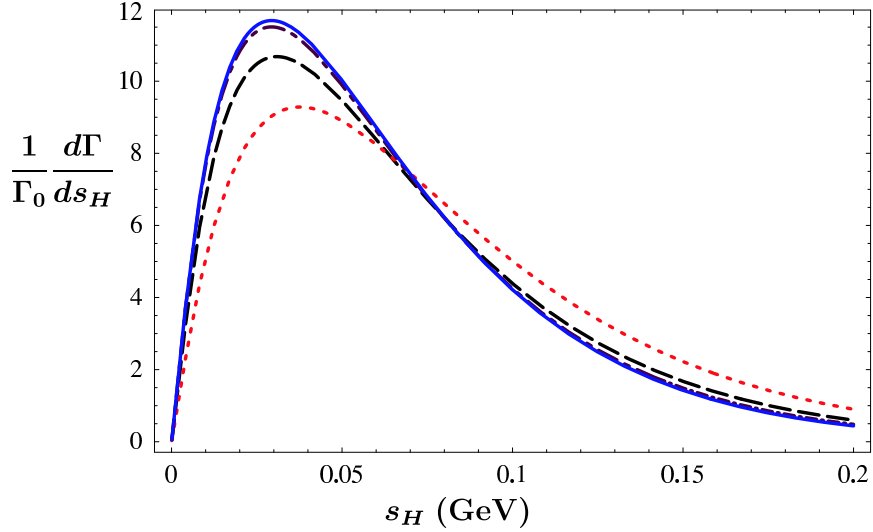}}
\caption{Plot of the kinematic corrections to the hadronic invariant mass
spectrum, 
\eqn{subleadingkin1}.  The dashed line is the leading order result
(\ref{leadingordersH}), 
while the solid line includes the full set of kinetic corrections.  The dotted
line corresponds 
to the expansion of the results of \cite{dFN} to subleading order, while the
dot-dashed 
line also includes the contribution from the $m_b^5$ term.  The difference
between 
the dot-dashed and solid curves is due to the expansion of the heavy quark
spinors.}
\label{kineticcorrections}
\end{figure}

The relative sizes of each of the terms in \eqn{subleadingkin2} is plotted in
\fig{kineticcorrections}, using 
the simple one-parameter  model for $f(\omega)$ introduced in
\cite{shapeleading2}
\begin{equation}\label{fmodel}
f_\mathrm{mod}(\omega) = \frac{32}{\pi^2 \Dlhat} (1+\omega/\Dlhat)^2
e^{-\frac4\pi(1+\omega/\Dlhat)^2}
	\theta(1+\omega/\Dlhat)
\end{equation}
and with $\Dlhat=0.1$.
Numerically, the most important of these corrections corresponds to smearing
the $m_b^5$ term, while the correction from expanding the quark fields is quite
small.

However, such large corrections may be misleading, since if they are universal
they may simply be absorbed in a redefinition of the leading 
order shape function.   Instead, one 
should look at the corresponding relation between the hadronic invariant mass
spectrum
and the $\bar{B} \to X_s\gamma$ photon energy spectrum.
One might expect that the effect of convoluting the $m_b^5$ term would cancel
in the relation,
 since both rates are
proportional to $m_b^5$.  However, in the $\bar{B} \to X_s\gamma$ spectrum only
three
powers of $m_b$ come from the kinematics, while two arise from the factor of
$m_b$ in the Wilson coefficient of $O_7$, and hence for this rate one should
only convolute three powers of $m_b$.  This may be verified by
writing the results of Ref.\ \cite{blm01} as
\begin{equation}\label{partialbsg}
{1\over \Gamma_0^s}{d\Gamma\over dx}(\bar{B} \to X_s\gamma)=f(1-x)-3(1-x)
f(1-x)+(1-x)
f(1-x)+\dots
\end{equation}
where once again the dots denote additional form factors, and the partonic
rate
is
\begin{equation}
\Gamma_0^s = \frac{G_F^2 |V_{tb}V_{ts}^*|^2 \alpha |C_7^{\mathrm{eff}} |^2
m_b^5}{32 \pi^4}.
\end{equation}
In the expression (\ref{partialbsg}), the second term corresponds to smearing
three powers of $m_b$ in the rate, while the third third term arises from the
expansion of the quark fields.  Thus, there is an incomplete cancellation of
the kinematic corrections between the two spectra.

\section{Phenomenology}
\label{phenomenology}

\subsection{The $\bar{B}\to X_u \ell \nu$ hadronic invariant mass spectrum and
the
$\bar{B} \to X_s \gamma$ photon energy spectrum}

As discussed in the previous section, there are large kinematic corrections
to the leading order results, largely due to the $m_b^5$ term in the rate.
However, these are reduced in the relation between the hadronic invariant
mass spectrum and the $\bar B\to X_s\gamma$ photon energy spectrum.  Similarly,
the T-product
$t(x)$ is universal for all processes involving $B$ meson decays (it only
differentiates between $B$ and $B^*$, $D$ and $D^*$ decays) and so its effects
similarly cancel.  Hence, it is useful to express the hadronic invariant mass
spectrum in terms of the experimentally measurable
$\bar B\to X_s\gamma$ photon energy spectrum.

The $\bar{B} \to X_s\gamma$ photon energy spectrum is given at tree level to
subleading order
in $1/m_b$ by \cite{blm01}
\begin{eqnarray} \label{photonenergyspectrum}
\frac{1}{\Gamma_0^s} \frac{d\Gamma}{d\hat{E}_\gamma} = 2F(1-2\hat{E}_\gamma)
\end{eqnarray}
where
\begin{eqnarray}
\label{Fdef}
F(x)=f(x) +
	\left[ h_1(x) + \frac{t(x)}{2} - 2 x f(x)
	- G_2(x) + H_2(x) \right] + \cdots.
\end{eqnarray}
(Note that at tree level only the operator $O_7$ contributes.  At one loop,
effects of other operators must be included \cite{mn01}).
Substituting this into \eqn{sHspectrum} gives
\begin{equation}\label{spectrumrelation}
{1\over\Gamma_0}{d\Gamma\over d\shhat}=\int_{0}^\infty \;d\omega \left\{
	(A(\omega,\shhat)+\delta A(\omega,\shhat)) F(\omega-\Dlhat)+\delta
F(\omega,\shhat,\Dlhat)\right\}
	d\omega
\end{equation}
where
$A(\omega,\shhat)$ and $\delta A(\omega,\shhat)$ are defined in
(\ref{leadingordersH}) 
and (\ref{subleadingkin2}), and
\begin{eqnarray}
\delta F(\omega,\shhat,\Dlhat)&=&\left[
	- \frac{2 \shhat ( 2 \shhat - 3 \omega)(\shhat-\omega)}{\omega^4}
G_2(\omega-\Dlhat) \right. \nn \\
	&\,& \hspace{0.5cm} + \left.
\frac{4\shhat(2\shhat+\omega)(\shhat-\omega)}{\omega^4} h_1(\omega-\Dlhat)
\right. \nn \\
	&\,&\hspace{0.5cm} + \left. \frac{2\shhat (2\shhat^2-4 \omega \shhat +
\omega^2)}{\omega^4} H_2(\omega-\Dlhat) \right]
\theta(\omega-\shhat)
\end{eqnarray}
contains the subleading shape functions.  (Note that the dependence on
the T-product $t(x)$ drops out of this relation.)

To extract $|V_{ub}|$, we are interested in the integrated rate
\begin{equation}
\Gamma_{s_H}(\shhat^c) = \int_0^{\shhat^c} \frac{d\Gamma}{d\shhat} d\shhat
\end{equation}
up to a maximum value $\shhat^c$.
The integrated rate for
$\btou$ is free of backgrounds from $\bar{B}\to X_c \ell \bar{\nu}_\ell$
for $s_H^c<{m}_D^2$, although because of experimental resolution the
experimental cut is
typically somewhat lower:  a recent \babar\ measurement \cite{babarsH} used
$s_H^c=(1.55\,\GeV)^2$.  From \eqn{spectrumrelation}, we have
\begin{equation}\label{intsHspectrum}
\frac{1}{\Gamma_0}\Gamma_{s_H}(\shhat^c)=\int_0^\infty\left\{(\tilde A(\omega, \shhat^c)+\delta
\tilde A(\omega, \shhat^c)) F(\omega-\Dlhat)+\delta\tilde F(\omega,
\shhat^c)\right\} d\omega
\end{equation}
where
\begin{eqnarray}
\tilde A(\omega, \shhat^c)&=&\theta(\shhat^c-\omega)+ \frac{
(\shhat^c)^3(2\omega-\shhat^c)}{\omega^4}
\theta(\omega-\shhat^c),\nonumber\\
\delta \tilde A(\omega, \shhat^c)&=&
\frac{8(\Dlhat-\omega)}{3}\theta(\shhat^c-\omega)+\frac{
2(\shhat^c)^2(\shhat^c(\Dlhat+5\omega)-9\omega^2)}
{3\omega^3}\theta(\omega-\shhat^c),\nonumber\\
\delta \tilde F(\omega,\shhat^c)&=&
-\frac{2}{3}\left(G_2(\omega-\Dlhat)+
2h_1(\omega-\Dlhat)+H_2(\omega-\Dlhat)\right)
\theta(\shhat^c-\omega)\nonumber\\
&&+\left[-\frac{(\shhat^c)^2(3(\shhat^c)^2-10\omega
\shhat^c+9\omega^2)}{3\omega^4} G_2(\omega-\Dlhat)\right.\nonumber\\
&&+\frac{2
(\shhat^c)^2(3(\shhat^c)^2-2\omega\shhat^c-
3\omega^2)}{3\omega^4}h_1(\omega-\Dlhat)\nonumber \\
&&+\left.\frac{(\shhat^c)^2(3(\shhat^c)^2-
8\omega\shhat^c+3\omega^2)}{3\omega^4}
H_2(\omega-\Dlhat)\right]\theta(\omega-\shhat^c).
\end{eqnarray}
Note that the upper limit of integration in $\omega$ corresponds to a photon
energy $x_\gamma=1+\Dlhat-\omega<0$; however, as discussed earlier, this region
is
outside the region of support of the shape function, and its contribution
should be
highly suppressed.  Thus, in the relation between the spectra we may set the
lower
limit on $x_\gamma$ to zero.

Comparing the two forms for the integrated spectrum $\Gamma_{s_H}(\shhat^c)$
in
(\ref{photonenergyspectrum}) and (\ref{intsHspectrum}), we can isolate the CKM
parameter $|V_{ub}|$:
\begin{eqnarray} \label{Vubextraction}
\frac{|V_{ub}|}{| V_{tb} V_{ts}^* |} &=& \left( 1 - \frac{1}{2}
\delta\Gamma_{s_H}(\shhat^c) \right)
	\left(\frac{6 \alpha |C_7^{\mathrm{eff}}|^2}{\pi}\right)^{1/2}\nonumber\\
	&&\times \left[\frac{\displaystyle\int_0^{\shhat^c}
\frac{d\Gamma}{d\shhat}\,d\shhat}
	{\displaystyle \int^{\hat{m}_B/2}_0 (\tilde A(\hat{m}_B-2\hat{E}_\gamma,
\shhat^c)+\delta
\tilde A(\hat{m}_B-2\hat{E}_\gamma, \shhat^c))\frac{d\Gamma}{d\hat{E}_\gamma} d\hat{E}_\gamma}
\right]^{1/2}
\end{eqnarray}
where we have defined
\begin{equation}\label{Gammaudefine}
\delta \Gamma_{s_H}(\shhat^c)={\displaystyle\int_0^\infty \delta
\tilde F(\omega,\shhat^c)\;d\omega\over \displaystyle\int_0^\infty \tilde
A(\omega,
\shhat^c) f(\omega-\Dlhat)\;d\omega}
\end{equation}
which contains the effects of the new subleading distribution functions.  For
comparison purposes, we also define
\begin{equation}\label{Gammaufulldefine}
\delta \Gamma_{s_H}^{\rm full}(\shhat^c)={\displaystyle\int_0^\infty \delta
\tilde A(\omega,\shhat^c) f(\omega-\Dlhat)+ \delta \tilde
F(\omega,\shhat^c)\;d\omega\over
\displaystyle\int_0^\infty \tilde A(\omega, \shhat^c)
f(\omega-\Dlhat)\;d\omega}
\end{equation}
which gives the full fractional subleading correction to the relation between
the two spectra.
To proceed further we must introduce a model for the shape functions.

\subsection{Shape function models}

The shape functions are nonperturbative functions which cannot at present be
calculated from first principles.  We do, however,
know several moments of these functions \cite{blm01}, and we can use this
information to constrain
possible models of the shape functions.

The leading order shape function is modeled with $f_\mathrm{mod}(\omega)$
defined in (\ref{fmodel}).  We will use three models of the subleading shape
functions.
The first was introduced in \cite{blm01}, based on the leading order function 
$f_\mathrm{mod}(\omega)$.  The subleading functions are defined as
\begin{eqnarray}
h_{1\;\mathrm{mod1}}(\omega) &=& \frac{\lambda_2}{m_b^2}
f'_\mathrm{mod}(\omega) \nn \\
G_{2\;\mathrm{mod1}}(\omega) &=& - \frac{2 \lambda_1}{3 m_b^2}
f'_\mathrm{mod}(\omega) \nn \\
H_{2\;\mathrm{mod1}}(\omega) &=& - \frac{\lambda_2}{m_b^2}
f'_\mathrm{mod}(\omega)  \nn \\
t_\mathrm{mod1}(\omega) &=& -\frac{\lambda_1 + 3 \lambda_2}{m_b^2}
f'_\mathrm{mod}(\omega)
\end{eqnarray}
to reproduce the leading terms in \eqn{fexp}.

The second model was introduced in \cite{mn02}, in which the subleading
functions are defined in terms of a single function
\begin{equation}
s(\omega, b) = -\frac{b^2}{\Dlhat^2} e^{-b (1+\omega/\Dlhat)}
	\left( b (1+\omega/\Dlhat) - 1 \right) \theta(1+\omega/\Dlhat).
\end{equation}
The dimensionless free parameter $b$ is constrained to be ${\cal O}(1)$
by the
requirement that the $n$th moments of the functions scale like $\Dlhat^{n+1}$.
We will take $b=1$ in our plots; larger values of $b$ reduce the effects of
the subleading shape functions.
We have
\begin{eqnarray}
h_{1\;\mathrm{mod2}}(\omega, b) &=& \frac{\lambda_2}{m_b^2} s(\omega,b) \nn \\
G_{2\;\mathrm{mod2}}(\omega, b) &=&- \frac{2 \lambda_1}{3 m_b^2} s(\omega,b)
\nn \\
H_{2\;\mathrm{mod2}}(\omega, b) &=&- \frac{\lambda_2}{m_b^2} s(\omega, b).
\nn
\\
t_\mathrm{mod2}(\omega, b) &=& -\frac{(\lambda_1 + 3 \lambda_2)}{m_b^2}
s(\omega, b).
\end{eqnarray}
Note that in the first model the subleading shape functions vanish at
$\omega=\Dlhat$, while in the second they are finite but nonzero.

\begin{figure}[tbp]
\centerline{\includegraphics[width=4.in]{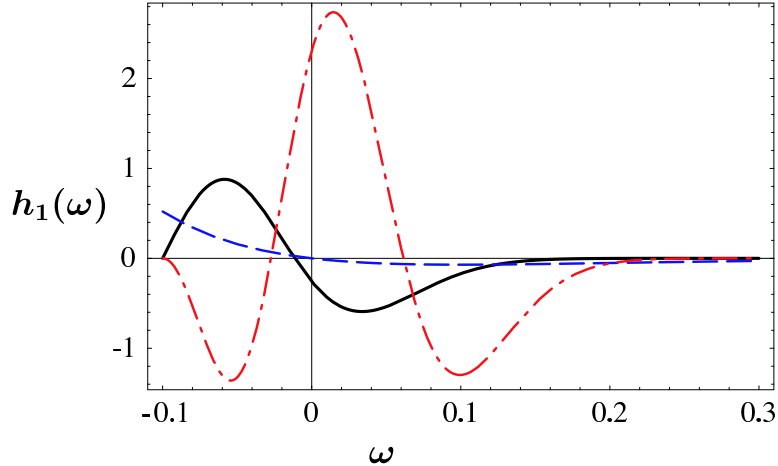}}
\caption{Three models of $h_1(\omega)$:  model 1 (solid curve), model 2
(dashed
curve) and model 3 (dot-dashed curve).}
\label{threemodels}
\end{figure}

In our third model\footnote{We thank C. Bauer for discussions of this model.},
we use a model for the subleading shape functions that has an additional sign
flip in the region of integration.  We take
\begin{eqnarray}\label{FfunctALT}
h_{1\;\mathrm{mod3}}(\omega) &=& \frac{\lambda_2}{m_b^2} f'_2(\omega) \nn \\
G_{2\;\mathrm{mod3}}(\omega) &=& - \frac{2 \lambda_1}{3 m_b^2} f'_2(\omega) \nn
\\
H_{2\;\mathrm{mod3}}(\omega) &=& - \frac{\lambda_2}{m_b^2} f'_2(\omega)
\nn
\\
t_\mathrm{mod3}(\omega) &=& -\frac{(\lambda_1 + 3 \lambda_2)}{m_b^2}
f'_2(\omega)
\end{eqnarray}
where
\begin{equation}\label{definefmodelb}
f_2(\omega)
=-\frac{32}{\pi^2\Dlhat}\left(\omega+\Dlhat\right)\theta
\left(1+\frac{\omega}{\Dlhat}\right) {d\over d\omega}
\left(\left(1+\frac{\omega}{\Dlhat}\right)^3
e^{-\frac4\pi(1+\omega/\Dlhat)^2}\right).
\end{equation}
We plot the function $h_1(\omega)$ in each of these models in
\fig{threemodels}.

Although the models have very different behaviour,  we can verify that they are
all
reasonable by calculating the first few moments ${\cal M}_n\equiv
\int_{-\infty}^\infty \omega^n H_2(\omega)d\omega$ in each model and showing
that they are of order $(\lqcd/m_b)^{n+1}$.  For $m_b=4.8$ GeV,  for model 1 we
find $|{\cal
M}_1|=(0.35\,\gev/m_b)^2, |{\cal M}_2|=0,\ |{\cal M}_3|=(0.35\,\gev/m_b)^4$ and
$|{\cal
M}_4|=(0.29\,\gev/m_b)^5$.  For model 2 we find $|{\cal
M}_1|=(0.35\,\gev/m_b)^2, |{\cal M}_2|=(0.49\,\gev/m_b)^3,\ |{\cal
M}_3|=(0.70\,\gev/m_b)^4$ and $|{\cal M}_4|=(0.90\,\gev/m_b)^5$, while for
model 3 the
corresponding moments are $|{\cal M}_1|=(0.35\,\gev/m_b)^2,\ |{\cal
M}_2|=(0.54\,\gev/m_b)^3,\ |{\cal M}_3|=(0.54\,\gev/m_b)^4$ and $|{\cal
M}_4|=(0.58\,\gev/m_b)^5$.  Thus, the first few moments of each model scale
like typical 
hadronic scales to the appropriate power.  Similar results are obtained for
the
moments of $G_2(\omega)$ and $h_1(\omega)$.

\subsection{Numerical results}

Both the Wilson coefficients and models for the shape functions depend on the
$b$ quark mass
$m_b$.   While in our formulas we are implicitly using the pole mass, it is
well-known that this leads
to badly behaved perturbative series, and so we expect that radiative
corrections to these results
will be minimized if a sensible short-distance mass is used instead.  The
$\overline{\mbox{MS}}$ mass
$\bar m_b(\bar m_b)$ is well-defined, but does not lead to small perturbative
corrections in $B$ decays \cite{lsw95, 
bsuv96}.
The ``threshold" masses, including the 1S mass, PS mass and kinetic mass, are
preferable in this context.  
At two-loops, a pole mass of $m_b=4.8$ GeV corresponds to a kinetic mass
$m_b^{\rm kin}(1\, \GeV)$ of 
about 4.6 GeV, PS and 1S masses of about $4.7\,\GeV$ and an
$\overline{\mbox{MS}}$ mass $\bar m_b(\bar 
m_b)$ of about 4.3 GeV. Thus, to give an estimate of the $m_b$ dependence of
our results, we plot them for 
$m_b=4.8$ GeV and $m_b=4.5$ GeV.

\begin{figure}[tbp]
\centerline{\includegraphics[width=6.5in]{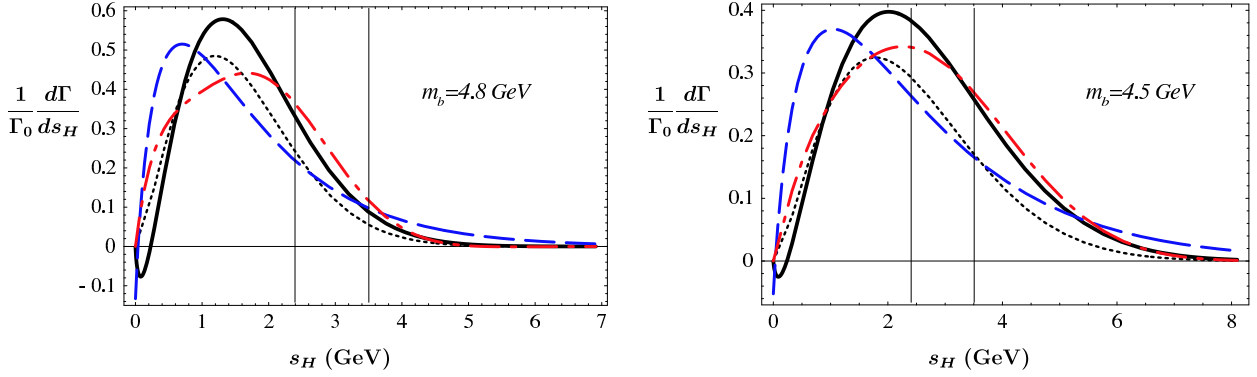}}
\caption{Model calculations of the hadronic invariant mass spectrum
$d\Gamma/ds_H$, for $m_b=4.8\,\GeV$ 
(a) and $m_b=4.5\,\GeV$ (b).  The dotted curve is the leading order result; the
other curves are the results in 
the models discussed in the text.  The curves are denoted as in
\fig{threemodels}.  The right vertical line 
denotes 
the kinematic limit $s_H=m_D^2$; the left line denotes the \babar\ cut
$s_H=(1.55\,\GeV)^2$.}
\label{twosHplots}
\end{figure}

\begin{figure}[tbp]
\centerline{\includegraphics[width=6.5in]{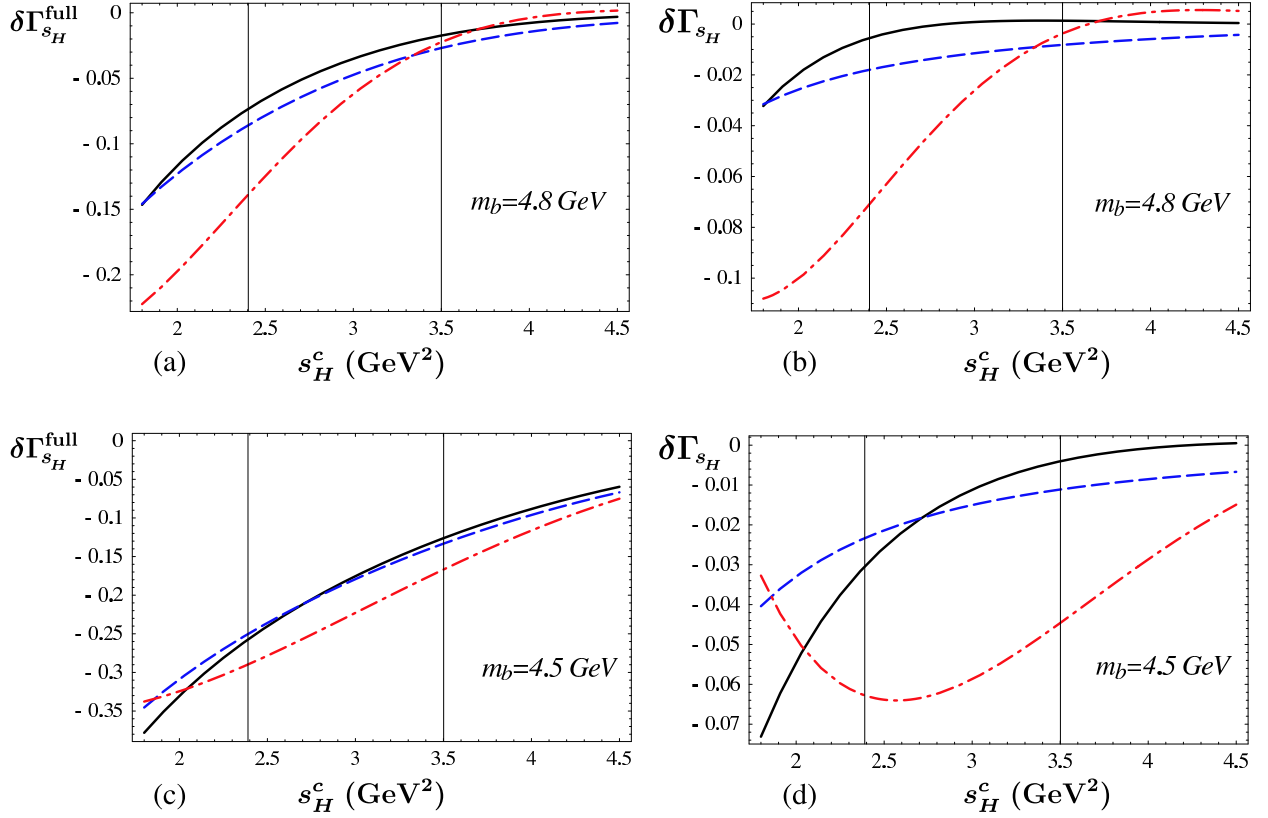}}
\caption{Model calculations of the fractional corrections
$\delta\Gamma_{s_H}^{\rm full}$ and 
$\delta\Gamma_{s_H}$ to the cut width, as defined in Eqs.\ (\ref{Gammaudefine})
and 
(\ref{Gammaufulldefine}), for $m_b=4.8\,\GeV$ (a,b) and $m_b=4.5\,\GeV$ (c,d).
The three curves refer to the three different models in \fig{threemodels}.  
$\delta \Gamma_{s_H}^{\rm full}$ includes all the subleading corrections,
including 
those proportional to the leading order shape function, while
$\delta\Gamma_{s_H}$ only 
includes the corrections from subleading shape functions.  The right vertical
line denotes 
the kinematic limit $s_H=m_D^2$; the left line denotes the \babar\ cut
$s_H=(1.55\,\GeV)^2$.}
\label{plotssH}
\end{figure}

In \fig{twosHplots}, we plot the hadronic invariant mass spectrum using the
three models 
of the previous section for the subleading corrections.  These corrections are
clearly large and model 
dependent over much of the spectrum.  
However, the integrated rate is much less sensitive to the subleading
corrections.  
The functions $\delta\Gamma_{s_H}(\shhat^c)$ and $\delta\Gamma_{s_H}^{\rm
full}(\shhat^c)$ defined
in Eqs.\ (\ref{Gammaudefine}) and (\ref{Gammaufulldefine}) are plotted in
\fig{plotssH} 
for the three models presented in the
previous section.  

{}From these figures it is clear that, at least for the particular models we
have chosen, the subleading shape functions do not contribute a large
uncertainty in the extraction of $|V_{ub}|$, and that the dominant subleading
effects are from the kinematic terms.   This should not be surprising:  since
there are no
$O(\lqcd/m_b)$ corrections to the total semileptonic decay rate \cite{operefs},
the
subleading corrections must vanish when integrated over the full spectrum.
Since
the experimental cuts include a large fraction of the rate, the contribution to
the 
integrated rate from the subleading corrections is correspondingly suppressed.
This
is evident from the plots in \fig{plotssH}, where the fractional correction
tends to zero as
the cut is increased.

\begin{figure}[tbp]
\centerline{\includegraphics[width=6.5in]{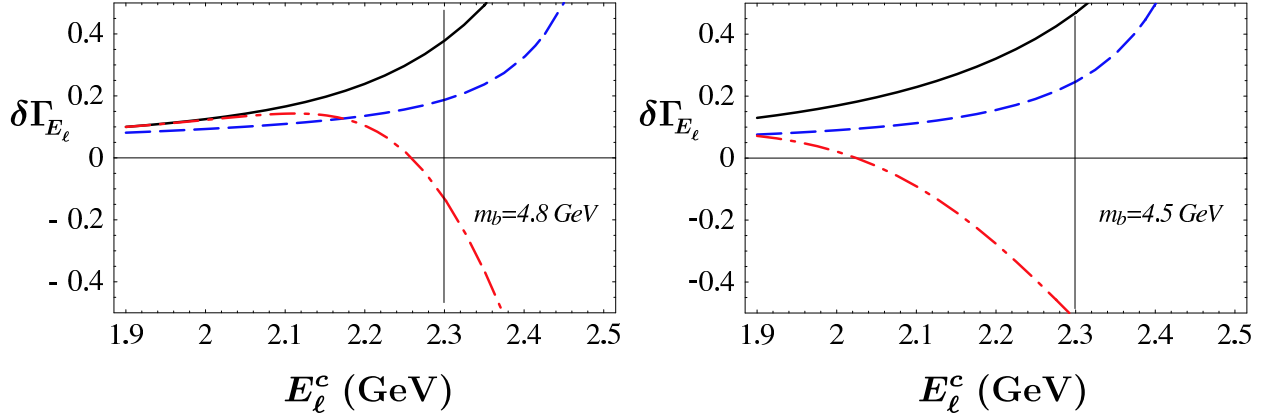}}
\caption{Model calculations of the fractional corrections
$\delta\Gamma_{E_\ell}$ for 
the semileptonic $B\to X_u$ decay width with a charged lepton energy cut, as 
defined in \eqn{Gammaelldefine}.  The three curves refer to the three different
models in \fig{threemodels}.  The solid line denotes the kinematic upper limit
from 
$B\to X_c$ decay.}
\label{plotsEe}
\end{figure}

It is useful to compare these results
with analogous results for the lepton energy spectrum in semileptonic $B$
decays,
given in \cite{blm02}.   In this case, only $\sim 10\%$ of the rate is
included, and the
subleading corrections are substantial.  The analogous
relation to \eqn{Vubextraction} is
\begin{equation}
\frac{|V_{ub}|}{| V_{tb} V_{ts}^* |} = \left( 1 - \frac{1}{2}
\delta\Gamma_{E_\ell}(\hat{E}_\ell^c) \right)
	\left( \frac{6 \alpha |C_7^{\mathrm{eff}}|^2}{\pi}\right)^{1/2}
\left[\frac{\int_{\hat{E}_\ell^c}^{\hat{m}_B/2}
	\frac{d\Gamma}{d\hat{E}_\ell}\, d\hat{E}_\ell}
	{\int_{\hat{E}_\ell^c}^{\hat{m}_B/2}
	8(\hat{E}_\gamma - \hat{E}_\ell^c)(1-\hat{E}_\gamma)\frac{d\Gamma}{d
\hat{E}_\gamma} \, d\hat{E}_\gamma} \right]^{1/2}
\end{equation}
where
\begin{equation}\label{Gammaelldefine}
\delta\Gamma_{E_\ell}(\hat{E}_\ell^c) =
2\frac{\int_0^{\hat{m}_B-2\hat{E}_\ell^c}
(\hat{m}_B-2\hat{E}_\ell^c-\omega)(H_2(\omega-\Dlhat)-h_1(\omega-\Dlhat)) \, d\omega}
	{\int_0^{\hat{m}_B-2\hat{E}_\ell^c}
        (\hat{m}_B-2\hat{E}_\ell^c-\omega)f(\omega-\Dlhat)\,d\omega}.
\end{equation}
In \fig{plotsEe} we plot $\delta\Gamma_{E_\ell}(E_\ell)$ for $m_b=4.8\
\GeV$ and $m_b=4.5\ \GeV$ in the three models used in this paper.
It is clear from the figures that for lepton cuts near the kinematic limit
$E_\ell=2.3$ GeV,
the uncertainty in $|V_{ub}|$ from higher order shape functions is much greater
for the lepton
energy spectrum than from the hadronic invariant mass spectrum.

\section{Conclusions}
\label{conclusions}

 We have calculated the hadronic invariant mass spectrum for
$\btou$ in terms of shape functions to subleading
order.
Introducing some simple models for the shape functions we have
studied the $d\Gamma/ds_H$ spectrum numerically.

Since we know little about the form of the subleading shape functions, it is
difficult to estimate the corresponding theoretical uncertainty in $|V_{ub}|$.
However, using the spread of models as a guide, we can conclude that the
largest
subleading effects are proportional to the leading order shape function, and
so, given
a determination of the shape function from $\bar B\to X_s\gamma$ decay, do not
increase the theoretical uncertainty.   Assuming our spread of
models provides a reasonable measure of the theoretical uncertainty, we can
conclude that the theoretical uncertainty in $|V_{ub}|$ due to higher order
shape
functions is at the few percent level.  This is substantially less than the
corresponding
uncertainty in the integrated lepton energy
spectrum with the current experimental cuts.  This is also much less than the
other sources
of experimental and theoretical error in the current measurements of the
integrated hadronic
energy spectrum.

\acknowledgments

We thank C.~Bauer and Z.~Ligeti for useful discussions.
This work is supported in part by the Natural Sciences and Engineering
Research Council of Canada.

\end{document}